\documentstyle[preprint, aps]{revtex}
\begin{document}

\draft

\title{
Suppression of Superconducting Critical Current Density by Small Flux Jumps in $MgB_2$ Thin Films}

\author{
Z. W. Zhao, S. L. Li, Y. M. Ni, H. P. Yang, Z. Y. Liu and H. H. Wen \cite{response}  
}
\address{
National Laboratory for Superconductivity,
Institute of Physics and Center for Condensed Matter Physics,
Chinese Academy of Sciences, P.O. Box 603, Beijing 100080, China\\
}

\author{
W. N. Kang, H. J. Kim, E. M. Choi, and S. I. Lee 
}
\address{
National Creative Research Initiative Center for Superconductivity and Department of Physics, Pohang University of Science and Technology, Pohang 790-784, Republic of Korea
}

\maketitle

\begin{abstract}
By doing magnetization measurements during magnetic field sweeps on thin films of the new superconductor $MgB_2$, it is found that in a low temperature and low field region small flux jumps are taking place. This effect strongly suppresses the central magnetization peak leading to reduced nominal superconducting critical current density at low temperatures. A borderline for this effect to occur is determined on the field-temperature ( H-T ) phase diagram. It is suggested that the small size of the flux jumps in films is due to the higher density of small defects and the relatively easy thermal diffusion in thin films in comparison with bulk samples. 
\end{abstract}


\section{Introduction}
For applications of superconductors, a high transition temperature $T_c$ and superconducting critical current density $j_c$ are desirable. The discovery of the new superconductor $MgB_2$ \cite{nagamatsu} with a remarkably high transition temperature of 39 K and high critical current density $j_c $ provides a promising candidate for such applications\cite{eom,perkins,wen1strongquan,kim,thompson}. One big issue concerns the stability of the critical current and what process dominates the flux motion in the new superconductor $MgB_2$. For bulk samples the flux dynamics and the vortex phase diagram have been intensively investigated \cite{finnemore,caplin,wen2phase}. Usually it is believed that the $j_c$ is controlled by either continuous flux creep ( in intermediate and high temperature region ) or sudden, large and discontinuous flux jumps ( in low temperature region )\cite{dou,zzw}. The former can give rise to a continuously and parallelly marching flux front and thus a continuous magnetization vs. field M(H) curve, whereas the latter will generate many big blasts at the flux front leading to big discontinuous steps on the M(H) curves. Since there are generally more defects ( so more pinning centers ) in thin films, it is interesting to know whether the same flux dynamics is occurring in thin films as in bulk samples. In this paper, we present the experimental observation of many small flux jumps ( SFJ ) in low temperature and field region of $MgB_2$ thin films. It is further shown that the central magnetization peak of the magnetization-hysteresis-loop ( MHL ) is smeared out and the nominal $j_c$ is suppressed by this effect in comparison to that due to the continuous flux creep.

\section{Experiment}
The thin films of $MgB_2$ were fabricated on (1102) $Al_2O_3$ substrates by using pulsed laser deposition technique, which was described clearly in Ref.\cite{kang}. They are typically 400 nm thick with predominant c-axis orientation ( c-axis is perpendicular to the film surface ). A rectangular sample of size $2.1mm \times 4.9 mm $ was chosen for the magnetic measurements. The temperature dependence of the diamagnetic moment was carried out by a Quantum Design superconducting quantum interference device ( SQUID, MPMS 5.5 T ) and the magnetization-hysteresis-loops ( MHL ) were measured with a vibrating sample magnetometer ( VSM 8T, Oxford 3001 ) at temperatures ranging from 2 K to $T_c$ and external field up to 8 T along the c-axis. The M(H) curve was measured with a field sweep rate of 0.01 T/s and integration time of 60 ms. The pressure of helium gas in the sample chamber for thermal exchange was kept at 0.04 bar during the measurement. 

\section{Results}
In the inset of Fig.1, we show the temperature dependence of the zero-field-cooled (ZFC) magnetization measursed by SQUID at $\mu_0$H = 0.001 T. It is clear that the superconducting transition temperature $T_c$ is about 38 K and the transition is rather sharp indicating a good quality of the film.

The main frame of Fig.1 shows a typical MHL measured at 2 K with the external field sweep rate of 0.01 T/s. One can see that the MHL is symmetric about the M = 0 axis showing the dominance of the bulk superconducting current here. Another interesting finding is that the MHL is almost closed at about 8 T, being very similar to the observation on bulk $MgB_2$\cite{wen2phase}. This relatively low irreversibility field $H_{irr}$ determined from the closing point of MHL was first attributed by Wen {\it et al.}\cite{wen1strongquan} to the existence of quantum vortex liquid in rather clean system of $MgB_2$. The nominal $j_c$ estimated using the Bean critical state model is about 1.2$\times 10^7 A/cm^2$ at 2 K and zero field which is about one order of magnitude higher than that in high-pressure synthesized bulk samples. This further indicates a rather good quality of the film. Worthy of noting here is that multiple small and irregular instabilities appear in the magnetization in a low field region. These irregular instabilities have magnitudes typically in the order of $10^{-3}$ emu. They are much higher than the noise background ( $10^{-5}$ emu ) of our VSM. The magnitude of these instabilities are much larger in low fields than in high fields ( where the M(H) curve becomes smooth ) although the field sweep rate and the data acquisition speed are the same indicating that the instabilities are not due to the noise background of the VSM. These instabilities have been rechecked with caution in a sense that after one week and two weeks, the same MHLs were obtained during the remeasuring process on $MgB_2$ thin films, confirming the reproducibility of the instabilities in $MgB_2$ thin films.

In order to know whether these irregular instabilities are related to the flux jumps appearing in bulk $MgB_2$ samples\cite{dou,zzw}, we have carried out detailed measurements at different temperatures. The results are shown in Fig.2. One can clearly see the following interesting findings: (i) The instabilities in the magnetization appear in a region below a certain value of field and temperature, e.g. below 1.3 T at 2 K; (ii) When these instabilities appear the central magnetization peak is strongly flattened out ( see e.g. the data of 2, 4, 6, and 8 K ) revealing a suppression of the nominal $j_c$; (iii) The magnetizations at 2, 4, 6 and 8 K nearly merge at low field, though at higher fields they are separated gradually showing the usual order of the $j_c$ vs. H and T, i.e., a higher $j_c$ at a lower field and temperature; (4) When the temperature is increased the small instabilities will evolve into some larger ones ( at 10 K ) and disappear completely at a higher temperature ( at 14 K ). The larger instabilities here look similar to those appearing in bulk samples, therefore it is tempting to regard these instabilities as small flux jumps. 

In Fig.3 we show a part of MHL measured at 4 K. One can clearly see that when the field is swept through 0.7 T from above, the continuous MHL curve becomes discontinuous: small and irregular instabilities appear in low field region. Interestingly these two regions are separated by a clear kink point on the MHL. The same feature appears for other MHLs measured at 2, 6 and 8 K. This kink may be understood in the following way: the SFJ appearing in low field region suppresses the magnetic moment ( thus the nominal critical current density ) of the sample; while in high field region the M(H) curve will behave in another way since no SFJ occurs there. However we are not sure whether this kink point is corresponding to a phase transition of the vortex system. 

\section{Discussion}
Fig.4 shows the field dependence of the nominal $j_c$ determined using the Bean critical state model via $j_c = 20 \Delta M/Va(1-a/3b)$, where $\Delta M$ is the width of the MHL, V, a and b are the volume, width and length ( $a < b$ ) of the sample, respectively. Although for system with flux jumps the Bean critical state model may be inapplicable, it can be used to estimate the nominal $j_c$ value for a qualitative comparison. At 14 K and near zero field, the estimated $j_c$ is as high as $ 1.7 \times 10^7 A/cm^2$, which is rather high. It is necessary to note that the $j_c$ value at 14 K is a real magnetic critical current density instead of a nominal one since here no flux jumps occur. However, at lower temperatures, the nominal $j_c$ is clearly suppressed by these SFJ in low field region. In the high field and temperature region the M(H) curve becomes continuous showing the gradual setting in of the normal flux creep. 

To get a more comprehensive understanding to the SFJ observed in a $MgB_2$ thin film, it is worthwhile to compare it with that in bulk $MgB_2$\cite{dou,zzw}. The MHLs of a high pressure synthesized bulk sample are presented in Fig.5. It is clear that after each flux jump,  the optimal slope may not fully recovered, while there is no suppression of the central magnetization peak ( and thus the nominal $j_c$ ) near zero field at 2 and 4 K in contrast to what observed in $MgB_2$ films. In other words, although strong flux jumps occur at 2 and 4 K in bulk, the outlines of the MHLs at these temperatures are still wider than those at higher temperatures where no flux jumps are observed.

Now we have a close look at the part of MHLs where the flux jumps occur.  In Fig.6(a) and Fig.6(b)  the magnetization M and the derivative dM/dH are plotted against the field in the flux jump region for the bulk and the thin film respectively. Two types of flux jumps can be clearly seen here for these two different samples. The jumps in the bulk sample are sparse and relatively large. After each jump, the magnetization changes significantly and then it gradually comes back to the main branch of MHL. A nearly constant field interval for every jump ( about 0.07 T ) is observed. During multiple measurements, the jumps repeatably tend to occur at the same field with the same magnitude. All these can be qualitatively understood based on the Swartz and Bean's adiabatic theory\cite{swartz}. The flux jump can be triggered when the gradient of magnetic flux profile inside the sample exceeds some critical value. After very short time ( usually in the order of $ms$ \cite{wertheimer} ) the magnetic profile drops to a new configuration with a lower gradient, then the flux only creeps slowly until the gradient of magnetic flux profile induced by varying the external field exceeds the critical value again. In each jump many vortices are involved in the thermomagnetic avalanche which normally expands to a large part of the sample volume. In thin films, however, the situation is completely different: many small local avalanches occur ( as shown in Fig.6(b)). Although the H-T region for flux jumps to appear doesn't change in different round of measurement of in $MgB_2$ thin films under identical conditions, the specific positions and the magnitude of the SFJ are however completely unrepeatable. All these cannot be explained by the adiabatic theory. The different avalanches observed in bulk samples and thin films may be induced by the different structural details and thermal diffusibility\cite{wipf}. For example, in bulk samples, there are many large grain boundaries, which act as strong pinning centers. The gradient of the flux profile near these boundaries can be broken at a certain limit. Once a blast occurs in a bulk sample, the thermal energy induced by drastic flux motion cannot easily diffuse out and be carried away by the environment. Therefore this self-heating will lead to increase the region in which the vortex instability occurs leading to a large jump on the magnetization in a bulk superconductor\cite{wertheimer}. One can also understand from this picture that the number of flux jumps cannot be large in bulk samples. In thin films the situation can be very different. On one hand scanning-electron-microscope ( SEM ) data indicates a high density of small defects formed during the preparation process of the thin films leading to much stronger critical current densities. Therefore there are many places for the avalanche to occur. On the other hand thermal diffusion is much easier in thin film samples due to their very small thickness and large surface area exposed to the environment. Therefore in thin films each avalanche is small in magnitude but the number of avalanches can be huge. This picture may give an explanation to many small vortex avalanches observed in Nb-film\cite{duran} and the YBCO film\cite{leiderer}. However why some of these small avalanches will grow in a dendritic structure\cite{duran} is still an open question. But clearly the very fine disorder structure and the relatively better thermal diffusion in thin films are two key factors to be considered here.

Finally we suggest a criterion to identify the specific region for these SFJ on the H-T vortex phase diagram. The boundary point for SFJ is defined as the clear kink point shown in Fig.3. Above this field no flux jumps could be observed above the noise background of the instrument. This is very helpful to illustrate the field and temperature region in which the application of the $MgB_2$ film can be hampered by the thermal instability. In Fig.7 a borderline $H_{SFJ}(T)$ for the SFJ is plotted together with the irreversibility line $H_{irr}(T)$ determined from the closing point of the MHL with a criterion of $\Delta M = 10^{-4} emu $. The SFJ appears only in the low temperature and low field region. Beyond this region the M(H) curves become continuously showing normal flux creep with very low creep rate\cite{wen3fluxfilm}.  Therefore it is safe to conclude that the application in large part of the vortex solid will not be influenced by the SFJ. However, it is important to point out that the SFJ observed in these films will prevent using SQUID device made from these thin films at low temperatures.

We have been aware of a recent result by Johansen {\it et al.} \cite{johansen} who found the SFJ and some dendritic avalanches in $MgB_2$ thin films by doing the magneto-optical measurement. One can see from their data that many tiny avalanches occur first at the edge of the film and some of them will gradually grow into a dendritic structure. Each tiny jump on our MHL may correspond to a local avalanche or the growth on one branch of the dendritic structure. These tiny jumps cannot be observed from the MHL measured by SQUID as presented by Johansen et al. since the SQUID has a very low speed for data acquisition. They can be seen clearly from our data measured by VSM with a fast data reading capacity. 
  
\section{Concluding remarks}
In conclusion, we have reported the observation of the suppression of the central magnetization peak and thus the nominal critical current density in the new superconductors $MgB_2$ film at low temperatures due to many small flux jumps. A comparison with a $MgB_2$ bulk is made. It is suggested that the small vortex avalanches in the thin film are closely related to the high density of small defects and the relatively easy thermal diffusion. A borderline for this effect to occur is determined on the H-T phase diagram. This gives helpful information for the application of $MgB_2$ films.

\acknowledgements
This work is supported by the National Science Foundation of China (NSFC 19825111) and the Ministry of Science and Technology of China ( project: NKBRSF-G1999064602 ). The work at Pohang University was supported by the Ministry of Science and Technology of Korea through the Creative Research Initiative Program.

\begin{figure}
\caption{
Magnetization hysteresis loop measured by VSM at 2 K. The curve is nearly closed at 8 T and many small magnetic instabilities can be seen at low field. The inset shows the temperature dependence of the ZFC magnetization of the $MgB_2$ film measursed by SQUID at $\mu_0$H = 0.001 T. The transition temperature $T_c$ is about 38 K and the transition is rather sharp indicating a good quality of the film. 
}
\end{figure}

\begin{figure}
\caption{
MHLs measured for a $MgB_2$ thin film at the temperatures of 2, 4, 6, 8, 10, 14, 18 and 22 K with the external field sweep rate 0.01 T/s. In low field region and at temperatures from 2 K to 10 K ( solid line ), there are many small flux jumps leading to the suppression of the nominal superconducting critical current density $j_c$.   
}
\end{figure}

\begin{figure}
\caption{
An enlarged view for the MHL at 4 K and the field ranging from 0 to 1.4 T on the field descending branch. We can clearly see that the SFJ begin at 0.7 T and a kink point appears on the MHL curve. The dotted line represents the extrapolation from the high field data. 
}
\end{figure}

\begin{figure}
\caption{
The nominal critical current density $j_c(H)$ derived from the magnetization in Fig. 2 based on the Bean critical state model. The maximal value reaches about $ 1.7 \times 10^7 A/cm^2$ at 14 K.
}
\end{figure}

\begin{figure}
\caption{
The MHLs measured for a high pressure synthesized $MgB_2$ bulk at temperatures of 2, 4, 6, 8 and 10 K with the field sweep rate 0.01 T/s. There is no suppression of the $j_c$ at 2 and 4 K. The flux jumps are relatively big and sparse.
}
\end{figure}

\begin{figure}
\caption{
Field dependence of the M(H) and dM/dH for (a) the bulk and (b) the film at 2 K and the same field region from 0 to 0.5 T on the field descending branch of MHL. For the bulk there are only few large flux jumps. And the occurrence of the jumps is repeatable even in detail. (b) For the film many small irregular jumps can be observed. The details of these small jumps are completely irreproducible.
}
\end{figure}

\begin{figure}
\caption{
The borderline $H_{SFJ}(T)$ ( filled circles ) separating the usual flux creep and the SFJ region.  The SFJ region is marked by the shaded area. The irreversibility line $H_{irr}(T)$ is represented by the filled squares. The lines are guide to the eye. The flux dynamics in the major part of vortex solid state is dominated by the elastic flux creep with very slow creep rate. 
}
\end{figure}

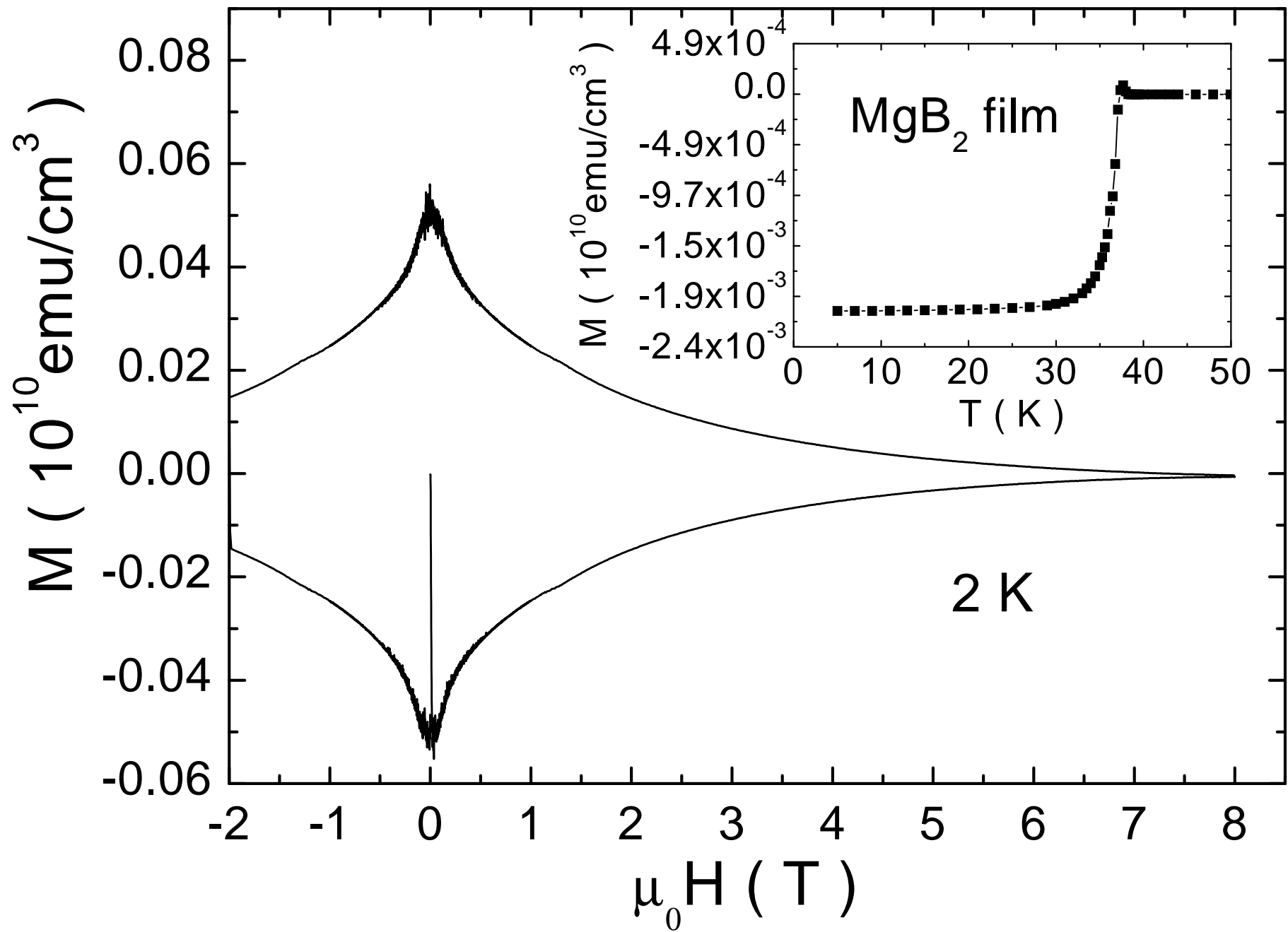
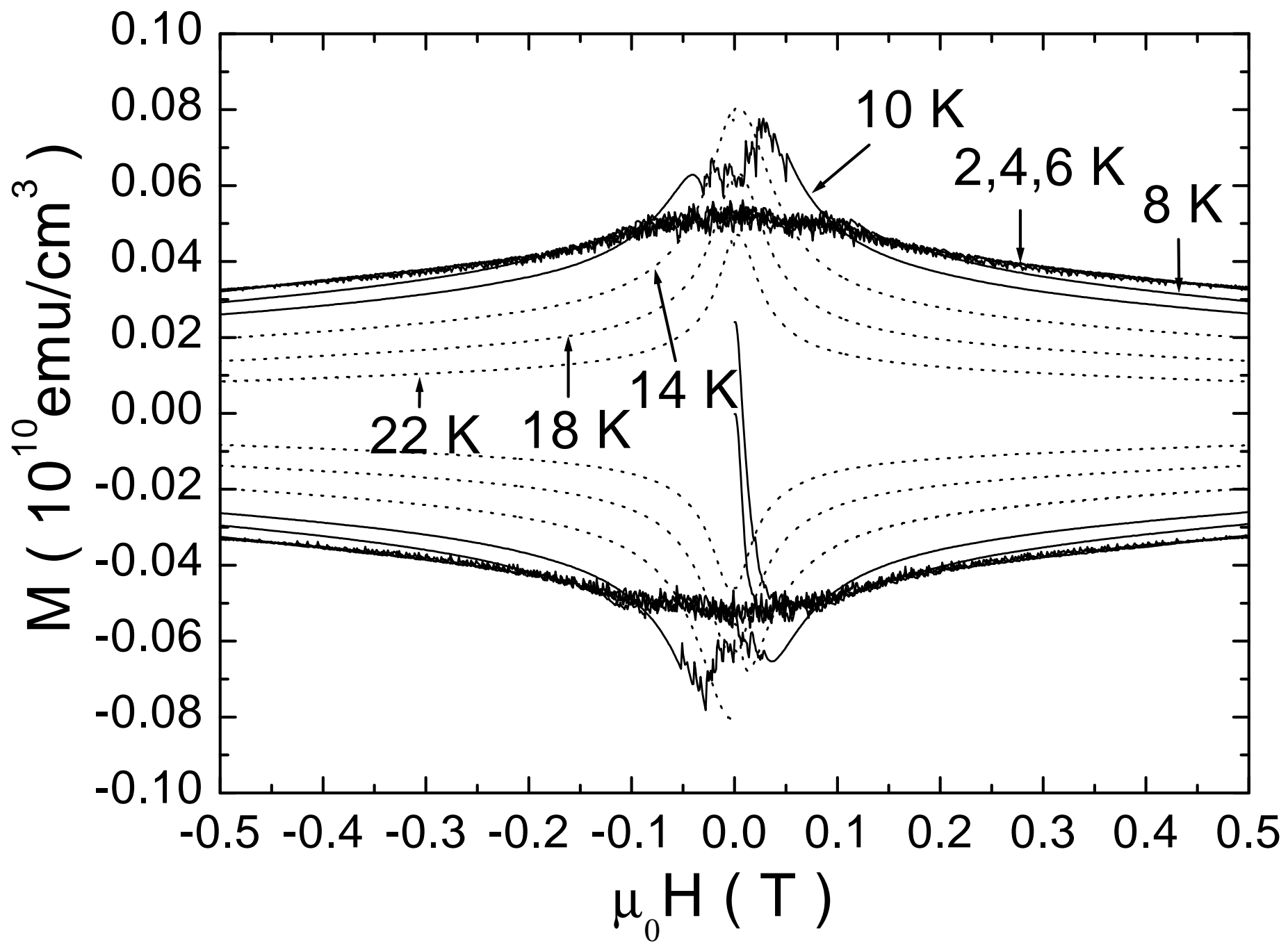
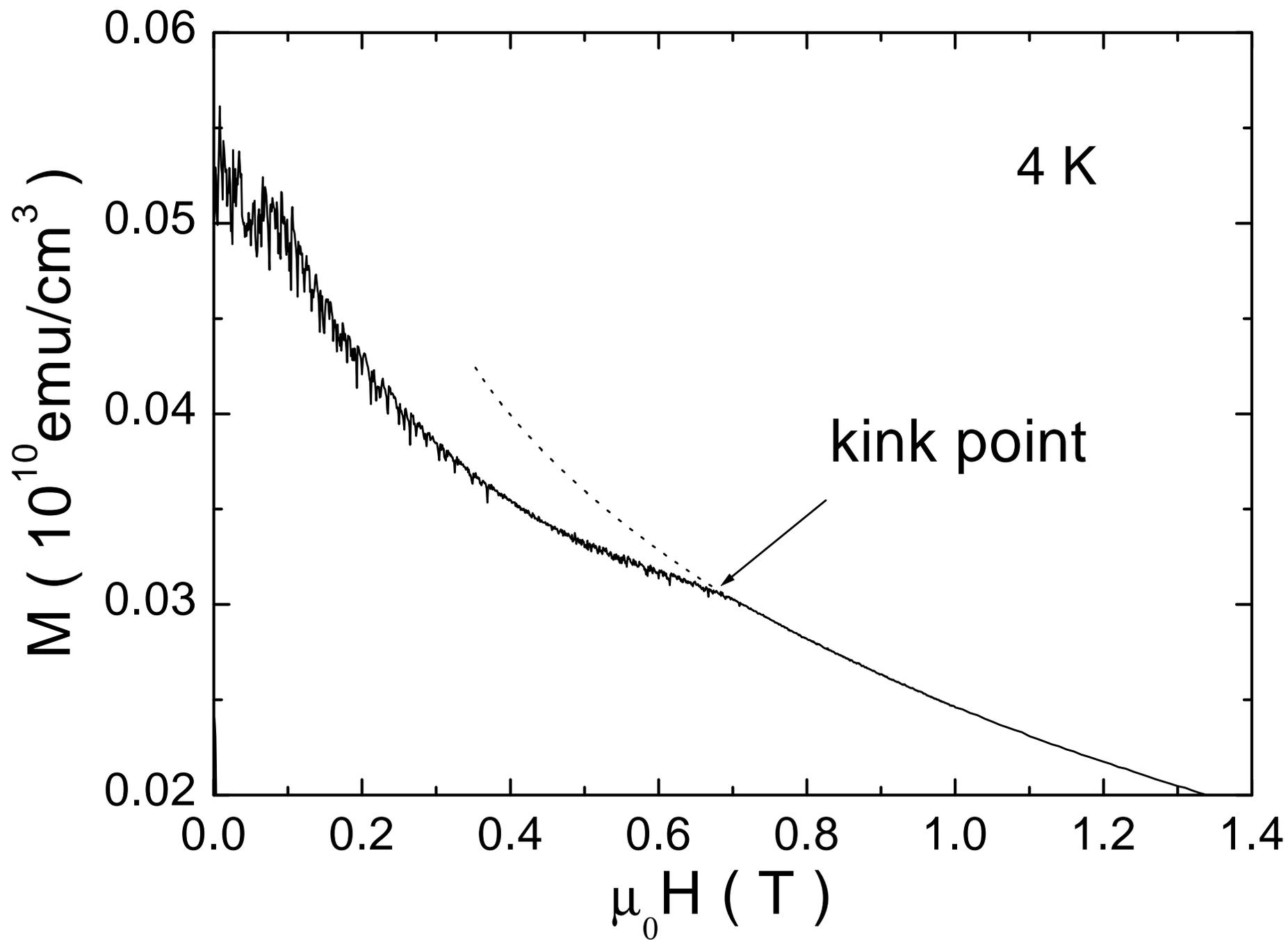
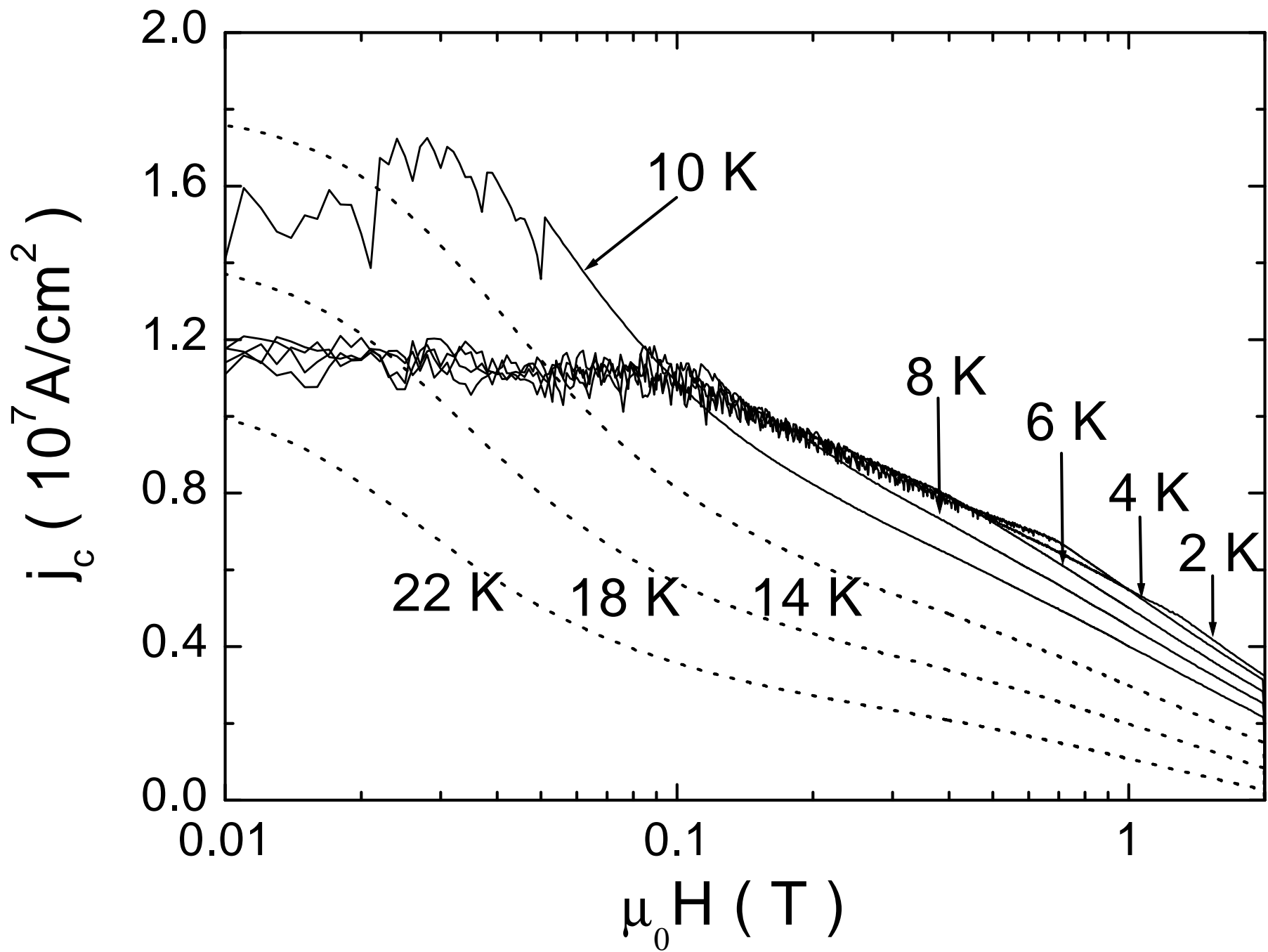
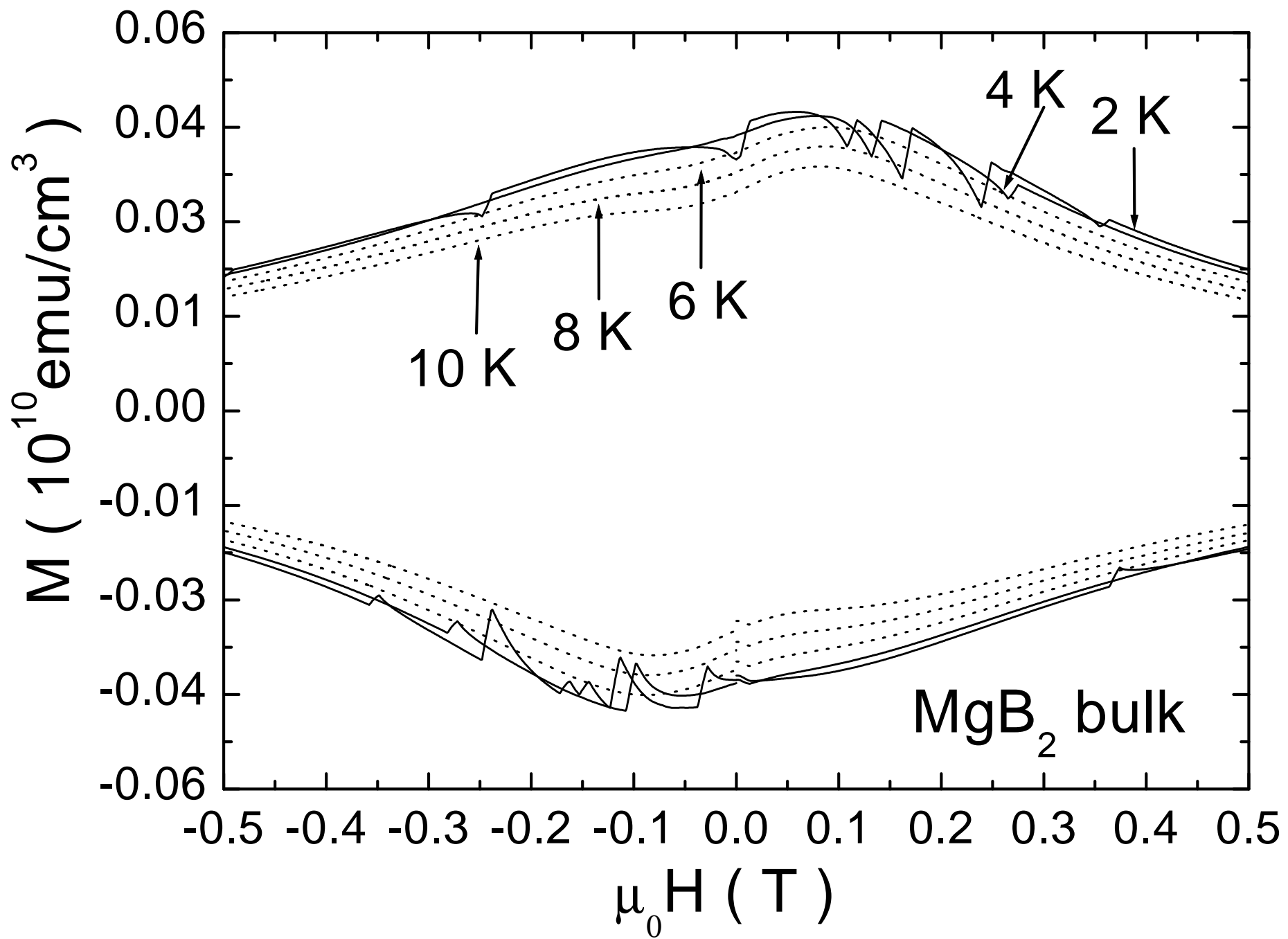
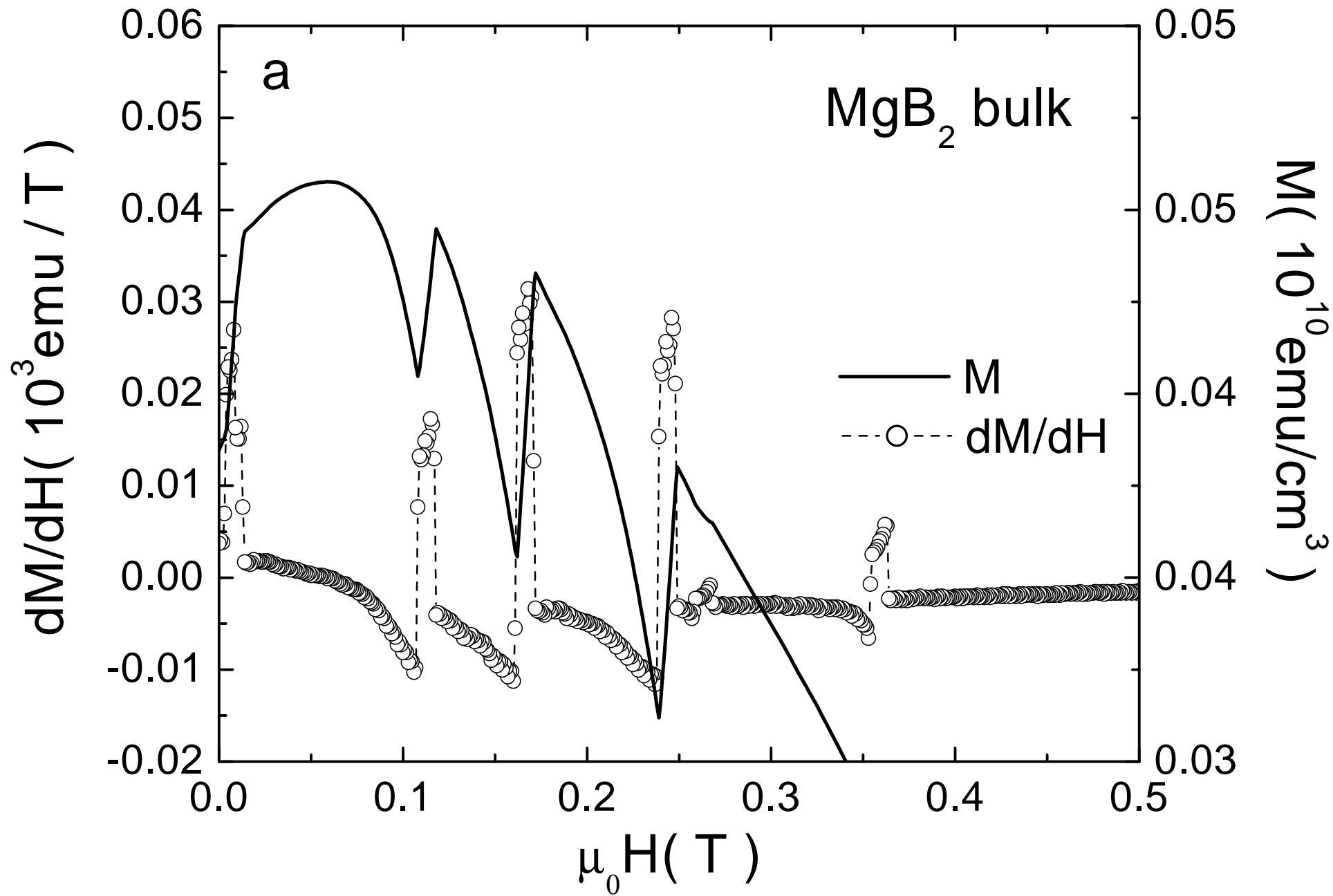
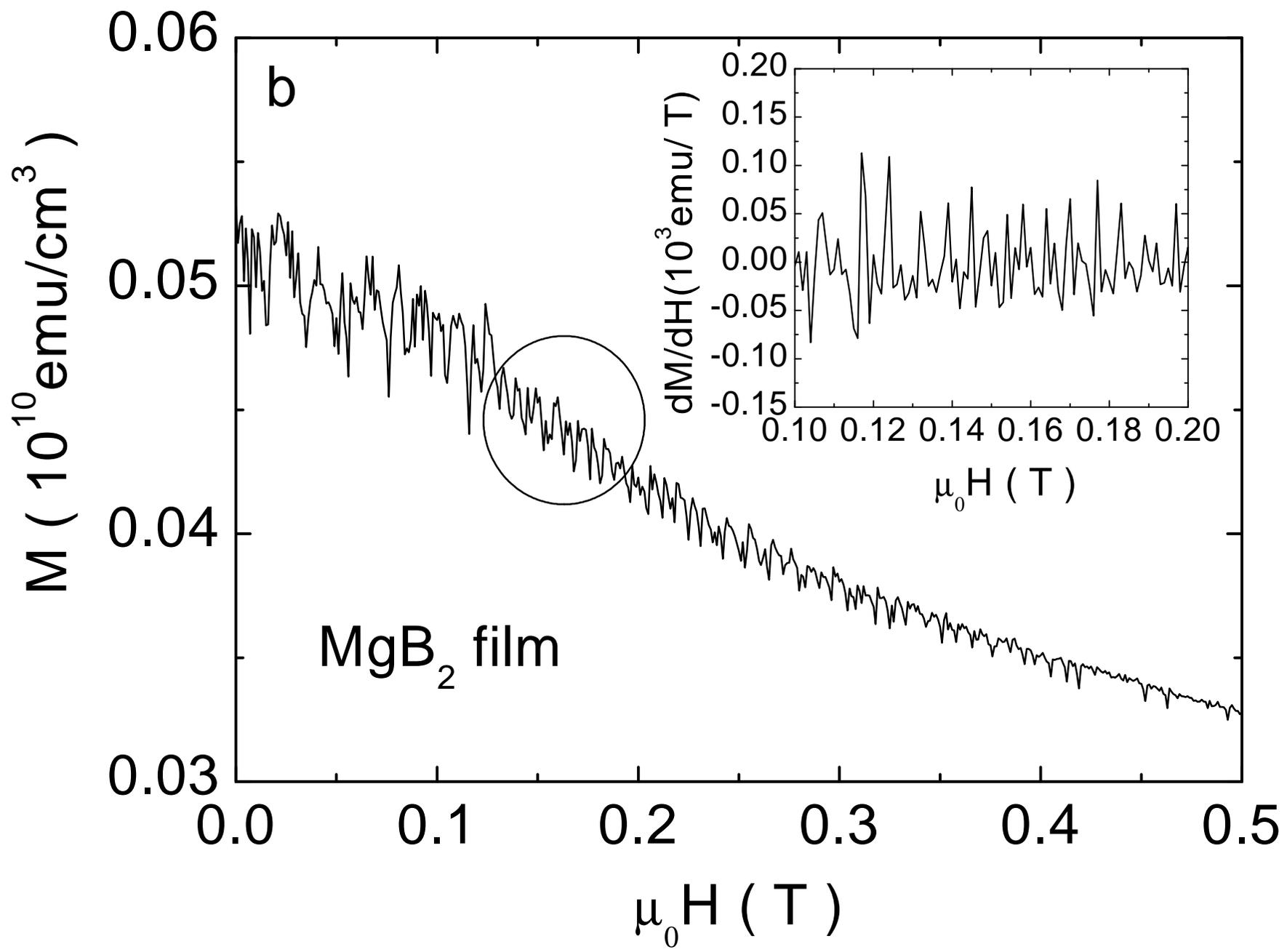
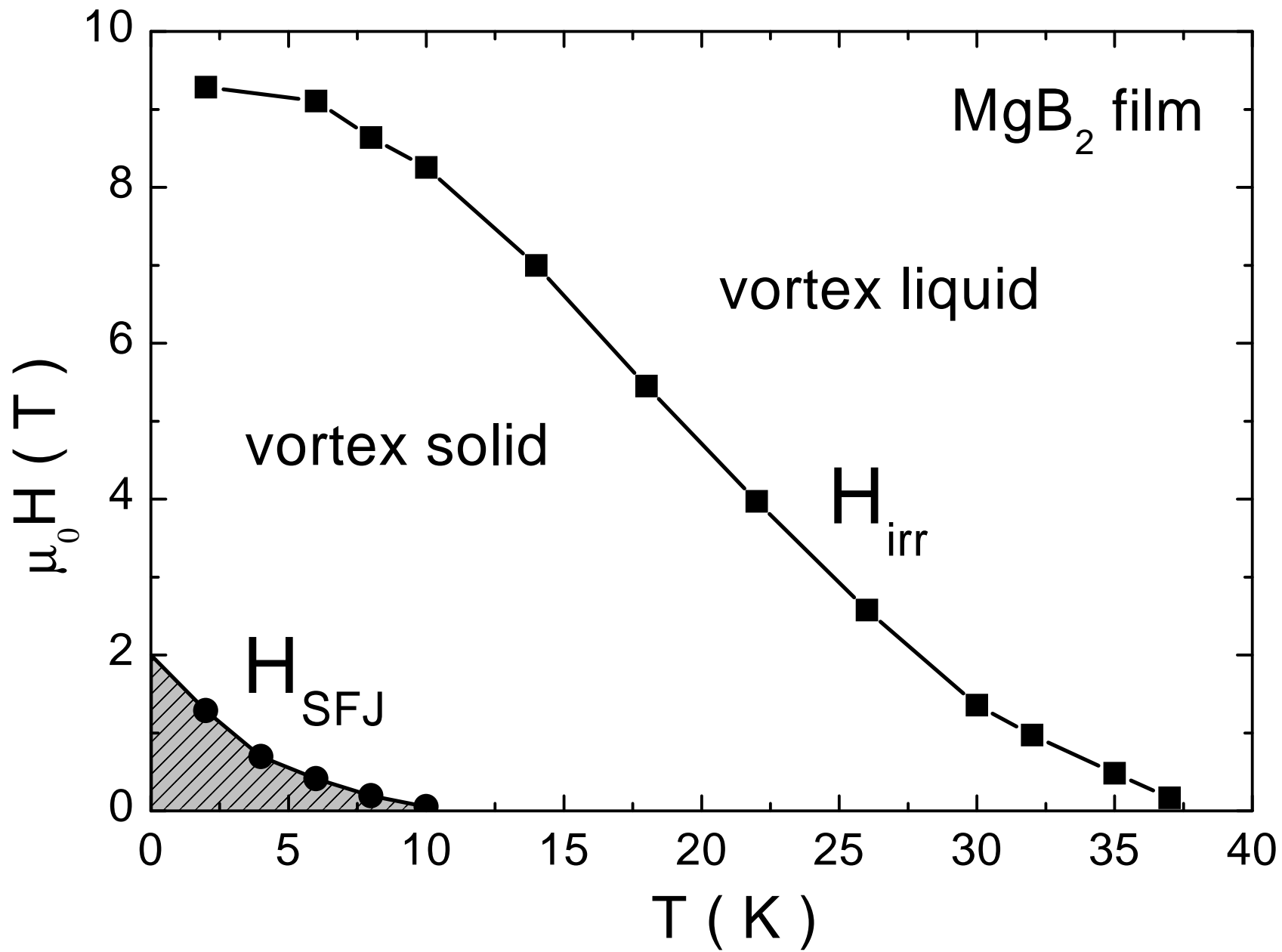
\end{document}